\RequirePackage{fix-cm}
\documentclass[twocolumn]{svjour3}          
\smartqed  
\usepackage{graphicx}
\begin{document}

\title{Intrinsic Multiple Andreev Reflections in layered Th-doped Sm$_{1-x}$Th$_x$OFeAs}

\titlerunning{Intrinsic Multiple Andreev Reflections in Sm$_{1-x}$Th$_x$OFeAs}

\author{T.E. Kuzmicheva        \and
        S.A. Kuzmichev \and
        S.N. Tchesnokov \and
        N.D. Zhigadlo
}

\institute{T.E. Kuzmicheva \at
              P.N. Lebedev Physical Institute of the RAS, 119991 Moscow, Russia \\
              Tel.: +7-499-132-60-48\\
              \email{kute@sci.lebedev.ru}           
           \and
           S.A. Kuzmichev, S.N. Tchesnokov \at
              M.V. Lomonosov Moscow State University, 119991 Moscow, Russia
              \and
              N.D. Zhigadlo \at
              ETH Zurich, CH-8093 Zurich, Switzerland
}

\date{Received: date / Accepted: date}

\maketitle

\begin{abstract}
Layered oxypnictide Sm$_{1-x}$Th$_x$OFeAs (Sm-1111) is an ideal candidate to be probed by intrinsic multiple Andreev reflections effect (IMARE) spectroscopy. Using the classical ``break-junction'' technique, we formed ballistic Andreev arrays of identical S--n--S-contacts, where $S$ is superconductor, and $n$ is a layer of normal metal. For $T < T_C$, the I(V) curve shows an excess-current and a subharmonic gap structure (SGS): a set of sharp dI(V)/dV-dips at positions which depend on the superconducting gap value, the number of junctions in the array, and the natural subharmonic order, thus manifesting the effect of intrinsic multiple Andreev reflections. Here we present the I(V) and dI(V)/dV with up to 4 SGS dips for Andreev arrays formed in optimally doped Sm-1111 with critical temperatures $T_C \approx 49$\,K, as well as in underdoped samples with $T_C \approx 37$\,K. We show that a number of Andreev subharmonics facilitates the determination of the superconducting gap with a better accuracy.

\end{abstract}

Sm$_{1-x}$Th$_x$OFeAs has a layered crystal structure typical for iron-based oxypnictides (1111 family) \cite{Kamihara} representing a stack of superconducting Fe--As blocks alternating with non-superconducting Sm(Th)--O spacers along the $c$-direction. Nonetheless, its electronic structure is quite different from that of other 1111-oxypnictides. Recent ARPES studies demonstrated \cite{Charnukha} that the Fermi surface of Sm-1111 comprises singular constructions formed by several band edges crossing the Fermi level.

In this study, we used polycrystalline Sm$_{1-x}$Th$_x$OFeAs samples with nearly optimal Th-doping and $T_C \approx 49$\,K, and underdoped samples with $T_C \approx 37$\,K. The high-quality polycrystallites possessing a single superconducting phase were synthesized under high pressure, as detailed in \cite{Zhigadlo,Zhigadlo2}. Superconducting properties of Sm-1111 samples were probed by intrinsic multiple Andreev reflections effect (IMARE) spectroscopy \cite{Pon_IMARE}, to use this method, we applied the ``break-junction'' technique \cite{Moreland}. The sample consisting of a thin rectangular plate of $4 \times 2 \times 0.2$\,mm$^3$ was attached to a spring sample holder by a liquid In-Ga alloy. Then the sample holder was cooled down to $T = 4.2$\,K, where a slight mechanical curving produced a cryogenic cleavage, thus creating two superconducting banks separated with a weak link (ScS-contact, $c$ -- constriction). In oxypnictides, the constriction usually acts as a thin normal metal, making it possible to observe multiple Andreev reflections in a ballistic SnS-contact \cite{Andreev,OTBK,Arnold,Kummel}. In a clean classical Andreev mode, it causes a current-voltage characteristic (CVC) with an excess current at low bias voltages (``foot'') and a subharmonic gap structure (SGS) in the dynamic conductance \cite{OTBK,Arnold,Kummel}. SGS presents a series of dI(V)/dV-minima at certain positions $V_n = 2\Delta/en, ~ n = 1,2, \dots$ The typical diameter of a ballistic SnS-contact in oxypnictides was estimated using Sharvin's formula: $a = 10 \div 30$\,nm \cite{Sharvin,Pon_LOFA}.

\begin{figure}
\includegraphics[width=0.5\textwidth]{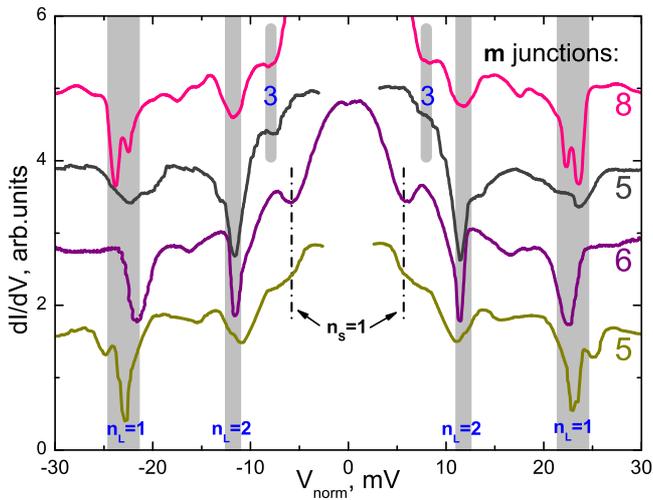}
\caption{Normalized dynamic conductance spectra of SnS-arrays (with $m =$ 5, 6, and 8 junctions in the stack) formed in different Sm$_{1-x}$Th$_x$OFeAs samples with nearly optimal critical temperatures $T_C^{bulk} = 49 \pm 2$\,K. $T = 4.2$\,K. Subharmonic gap structure for the large gap $\Delta_L = 11.8 \pm 1.2$\,meV is highlighted by grey areas and $n_L$ labels. Peculiarities of the small gap $\Delta_S = 2.8 \pm 0.3$\,meV are marked by dash-dot lines and $n_S$ label.}
\end{figure}

 A layered sample exfoliates along the $ab$-planes with the formation of steps and terraces along the $c$-direction. Steps could be realized as stack structures of S--n--S--n-- \dots --S-type. Considering such array as a sequence of $m$ identical SnS-junctions (formed in a stack of superconducting Fe--As blocks and Sm(Th)--O spacers), it becomes obvious that the position of dI(V)/dV-peculiarities related to IMARE will scale with $m$:

\begin{equation}
V_n = \frac{2\Delta}{en} \times m, ~~~~~ n,m = 1,2\dots
\end{equation}

This is the case for all quasi-two-dimensional superconductors studied. The break-junction CVC often demonstrates the subharmonic gap structure at bias voltages several times higher than the expected $2\Delta/en$ value \cite{Pon_IMARE,EPL}. Remarkably, IMARE is well-observed even in polycrystalline samples: if crystallites are larger than the contact diameter $a$, and intergrain solidity is stronger than interlayer one (the latter is typical for annealed layered samples), steps and terraces appear on the surface of cracked crystallites \cite{EPL}. Strictly speaking, IMARE is similar to intrinsic Josephson effect \cite{Nakamura}, firstly observed in cuprates \cite{Pon_IJE}.

\begin{figure}
\includegraphics[width=0.5\textwidth]{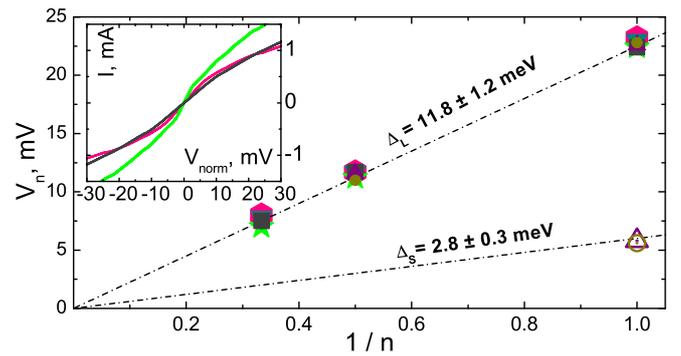}
\caption{The dependence of Andreev minima positions versus their inverse number, $1/n$, for the spectra shown in Fig. 1. Solid symbols depict the large gap dips (their color corresponds to the curves in Fig. 1), open symbols relate to the small gap. The inset shows typical current-voltage characteristics for SnS-Andreev arrays.}
\end{figure}

The first data of IMARE spectroscopy in nearly optimal Th-doped Sm-1111 ($T_C = 49 \pm 2$\,K) are presented in Figs. 1, 2. Since the dynamic conductance in an array scales in accordance with the formula (1), to determine the gap value one should divide the bias voltages obtained from the raw dI(V)/dV spectrum by the corresponding natural number of junctions in the stack $m$. Such normalization was done for the dynamic conductance spectra of several Andreev arrays with various $m = 8, 5, 6, 5$, respectively (see Fig .1). The resulting spectra correspond to single SnS-junctions. Although the arrays were formed in different Sm-1111 samples with similar $T_C$, the rather good coincidence between the positions of main gap peculiarities is obvious. The sharp minima marked by grey vertical areas, are rather symmetrically located at $V_1 \approx \pm 23$\,mV ($n_L=1$), $V_2 \approx \pm 11.8$\,mV ($n_L=2$), and $V_3 \approx \pm 7.8$\,mV ($n_L=3$), and satisfy formula (1), thus forming SGS for the large gap. The presence of a reproducible fine structure within the range $15 \div 17$\,mV would be an issue of further studies. Clearly, the dependence between the Andreev dip positions and their inverse number, $1/n$, is a line crossing the origin. Taking the average slope of this line, one can easily determine the large gap value $\Delta_L \approx 11.8$\,meV (see Fig. 2) with no need for any additional fitting of the dI(V)/dV even at $T \rightarrow T_C$ \cite{Kummel}. At the same time, CVC for such arrays demonstrates a pronounced ``foot'' (an excess current area at low bias voltages) typical for classical high-transparency Andreev mode \cite{Kummel}. The resistance of the arrays under study, $R = 20 \div 30 \rm{\Omega}$, in accordance with Sharvin's formula estimations \cite{Sharvin,EPL}, is large enough for ballistic transport feasibility, and, therefore, for observation of IMARE.

 \begin{figure}
\includegraphics[width=0.5\textwidth]{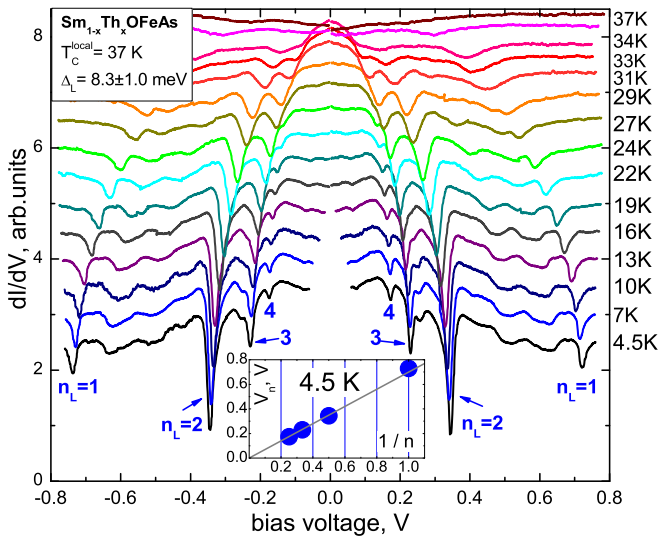}
\caption{Raw dynamic conductance spectrum for SnS-array (of $m = 42 \pm 3$ contacts) in overdoped Sm$_{1-x}$Th$_x$OFeAs sample with $T_C^{local} \approx 37$\,K. The $n_L$ labels point to the positions of Andreev subharmonics. The spectra were shifted vertically with the temperature increasing for clarity. On the contrary, the absolute value of the dynamic conductance decreases with $T$ increasing. The inset shows the linear dependence of subharmonic gap structure positions on $1/n$ at 4.5\,K.}
\end{figure}

We determined the corresponding BCS-ratio $2\Delta_L/k_BT_C \approx 5.6$. It exceeds the standard weak-coupling limit 3.52, at the same time, however, it is similar to that in optimal GdO(F)FeAs \cite{EPL,JPCS} or in Sm(Th)-1111 samples with $T_C = 37 \div 45$\,K \cite{Sm_JETPL}. As for the small gap in Sm-1111, the intensity of its SGS is evanescent, thus preventing the experimental observation of $\Delta_S$. The reason could be understood according to Sharvin's formula extended to the two-band case, for which the total conductance of ballistic contact is $e^2 a^2 (n/p_F^e + p/p_F^h)$, where $n$ and $p$ are the concentrations of electrons and holes, respectively. Such two-band simplification could be used directly, due to a unique property of nearly optimally-doped superconductors of the 1111 family described in \cite{Charnukha,Charnukha_disc}. Since the average Fermi momentum for electrons is nearly preserved for distinct doping level, the main difference in channels conductivity is determined by the ratio of electron and hole concentrations. Due to electron band flattening in the vicinity of the Fermi level, $n$ is small, and the hole band (driving in 1111 compounds the large order parameter \cite{Charnukha,Singh,Kondo}) dominates, hence suppressing any peculiarities in the tunneling spectra caused by the small order parameter. However, in the best contacts, the small gap occasionally manifests itself as a strongly smeared but reproducible first-order Andreev dip at $V \approx \pm 5.6$\,mV (see lower spectra in Fig. 1), leading to $\Delta_S \approx 2.8$\,meV. The latter value is also close to the small gap in Gd-based oxypnictides with the same $T_C$ \cite{EPL,JPCS}.

IMARE is well-observed also in underdoped Sm$_{1-x}$Th$_x$OFeAs (see Figs. 3, 4). In Fig. 3 we present the raw dynamic conductance spectrum for an SnS-array in the temperature range from 4.5\,K (lower curve) to $T_C = 37$\,K (upper curve). For clarity, the spectra were shifted along the vertical axis with increasing temperature. We emphasize that the absolute value of these dynamic conductance spectra decreases as $T$ increases. Noteworthy, the dI(V)/dV data look very similar to the dynamic conductance for the Andreev array in MgB$_2$ (see Fig. 1 in \cite{SSC}). At $T=4.5$\,K we observe four sharp dips (marked by $n_L$ labels) corresponding to IMARE in $\Delta_L$ band. While heating, the minima gradually move towards the zero bias becoming less intense, until they vanish at the local critical temperature of the contact $T_C^{local} = 37$\,K. The linearization of the dI(V)/dV-spectrum points to the contact area transition to a normal state. The position of the main conductance dips depends linearly on the inverse subharmonic order, as shown in the inset of Fig. 3, therefore, the dips comprise the SGS of the large gap. Earlier we demonstrated both the scaling of the dip positions with the corresponding natural numbers $m$, as well as the sharpening of the dips with increasing $m$ \cite{EPL}. Note, these features of array contacts cannot be explained by model of a sequence of intergrain contacts. On the contrary, the experiment proves that these peculiarities relate to bulk effects. The temperature behaviour of the $V_n$ position of the first (circles), second (squares), third (triangles), and fourth (rhombs) conductance dip plotted in Fig. 4 is similar, which also indicates that the observed peculiarities display properties related to the same condensate.

To determine accurately the number of junctions in the stack of a large $m$ is rather difficult but possible if several Andreev subharmonics are visible, and an unambiguous set of statistics is available. Since this is the case, we could estimate $m$. Given the average gap value from the $V_n(1/n)$ dependence (the inset of Fig. 3), $\Delta_L \times m \approx 350$\,meV, the corresponding scaled BCS-ratio $2\Delta_L \times m / k_BT_C^{local} \approx 219.5$, and the range of BCS-ratios obtained earlier in Sm(Th)-1111 $2\Delta_L/k_BT_C^{local} = 5.2 \pm 0.3$ \cite{Sm_JETPL,UFN}, we could estimate $m = 42 \pm 3$ junctions in the array, and for $\Delta_L = 8.3 \pm 0.6$\,meV the large gap value. Note that the uncertainty due the number of layers here is less than 10 \% of the value that we typically obtain with our experimental setup. Simple averaging over all the $V_n(T)$ and a division by $m=42$ yields the approximate temperature dependence of the large gap (see inset of Fig. 4, circles).

\begin{figure}
\includegraphics[width=0.5\textwidth]{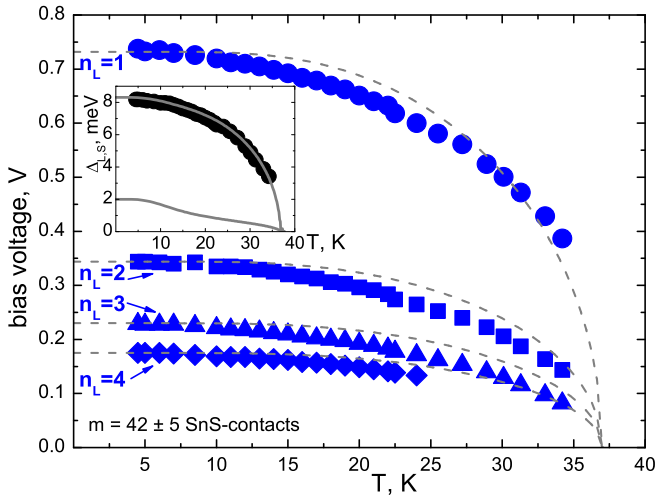}
\caption{The positions of the first (circles), second (squares), third (triangles), and fourth (rhombs) Andreev dips versus temperature for the dynamic conductance spectra from Fig. 3. $T_C^{local} \approx 37$\,K. The large gap value $\Delta_L = 8.3 \pm 0.6$\,meV. Dashed lines correspond to single-gap BCS-like dependences. The inset shows temperature dependence of the large gap (circles) obtained experimentally, and the theoretical $\Delta_{L,S}(T)$ approximation (bold lines).}
\end{figure}

Clearly, all the temperature dependences shown in Fig. 4 lie slightly lower than standard single-band BCS-type (depicted by dashed lines). This deviation is typical for all 1111 we studied \cite{JPCS,Sm_JETPL,UFN}, as well as for other multigap superconductors \cite{SSC,Pon_MgB2,Ba,KNa,Li,Gd_Superstripes,FPS}, and may be caused by the presence of a second condensate with a small gap, and a nonzero interband coupling ($k$-space proximity effect \cite{Yanson}). The gap temperature dependence can be fully described by a system of gap equations (with renormalized BCS-integral) by Moskalenko and Suhl \cite{Mosk1,Suhl,fit} comprising a $2\times2$ matrix of coupling constants $\lambda_{ij}$. In 1111-oxypnictides, we observe a scaling of both the superconducting gaps and $\lambda_{ij}$ within the wide range of critical temperatures $T_C = 21 \div 53$\,K \cite{UFN}. For a rough estimate, we could also refer to our statistics to restore the temperature dependence of the small gap. Firstly, taking into account the gap ratio $\Delta_L/\Delta_S \approx 4.2$ using the data of Fig. 1, and assuming it constant \cite{UFN}, we obtain the expected value of the small gap $\Delta_S \approx 2$\,meV. Secondly, by considering the earlier data on relative $\lambda$ values in Sm-1111 \cite{Sm_JETPL,UFN}, $\lambda_{LL}:\lambda_{SS}:\lambda_{LS}:\lambda_{SL} = 1:0.7:0.2:0.02$, it is possible to calculate the approximate $\Delta_{L,S}(T)$, which are shown in the inset of Fig. 4 as bold lines.

Note that such approximate calculations show an excellent agreement between the experimental and theoretical $\Delta_L(T)$. The expected behaviour of the small gap differs strongly from the BCS-type. While the temperature increases, $\Delta_S(T)$ first decreases significantly, then slowly fades untill the local critical temperature is reached. Such initial decrease of the small gap causes the $\Delta_L(T)$ bending in relation to the BCS-type dependence. Generally, in accordance with the theory \cite{Mosk1,Suhl,fit}, the higher the density of states in the ``driven'' band, the stronger the bends observed in the $\Delta_L(T)$ dependence \cite{UFN,fit}. Given the experimental $\Delta_L(T)$, we may conclude that Sm-1111 shows a ``driven'' condensate with a small gap, with significant Fermi-level density of states over $T_C$. Returning now to singular electron Fermi surfaces in Sm-1111 observed recently \cite{Charnukha}, the electron bands have: a) low concentration of charge carriers (since the electron bands barely cross the Fermi level), but b) significant density of states $N_S$.

In conclusion, we presented a pioneer study of the optimally Th-doped Sm$_{1-x}$Th$_x$OFeAs system with $T_C \approx 49$\,K, as well as underdoped Sm(Th)-1111 samples with $T_C \approx 37$\,K by intrinsic multiple Andreev reflections effect (IMARE) spectroscopy. IMARE is reproducibly observed in the Andreev arrays under study (with up to $\sim 42$ contacts in a sequence), causing the presence of excess current in I(V)-characteristics and subharmonic gap structure (SGS) in the dynamic conductance. The pronounced SGS facilitates the accurate determination of the large gap value, up to $\Delta_L \approx 11.8$\,meV at maximal critical temperatures, and the measurement of its temperature dependence. The small gap is less pronounced in the IMARE experiment which, however, is in a good agreement with ARPES stuides \cite{Charnukha}.

\begin{acknowledgements}
We thank Ya.G. Ponomarev, A. Bianconi, A. Charnukha, and A. Boris for useful discussions. The work was partially supported by RFBR grants 13-05-01451a and 13-02-01180a.
\end{acknowledgements}

\end{document}